# The Economics of Coal Phaseouts:
## Auctions as a Novel Policy Instrument for the Energy Transition

By Sugandha Srivastav* and Michael Zaehringer

*sugandha.srivastav@smithschool.ox.ac.uk

**Abstract**

The combustion of coal, the world's most polluting form of energy, must be significantly curtailed to limit global average temperature increase to well below 2°C. The effectiveness of carbon pricing is frequently undermined by sub-optimally low prices and rigid market structures. Consequently, alternative approaches such as compensation for the early closure of coal-fired power plants are being considered. While bilateral negotiations can lead to excessive compensation due to asymmetric information, a competitive auction can discover the true cost of closure and help allocate funds more efficiently and transparently. Since Germany is the only country till date to have implemented a coal phaseout auction, we use it to analyse the merits and demerits of the policy, drawing comparisons with other countries that have phased out coal through other means. Germany's experience with coal phaseout auctions illustrates the necessity of considering additionality and interaction with existing climate policies, managing dynamic incentives, and evaluating impacts on security of supply. While theoretically auctions have attractive properties, in practice, their design must address these concerns to unlock the full benefits. Where auctions are not appropriate due to a concentration in coal plant ownership, alternative strategies include enhanced incentives for scrappage and repurposing of coal assets.

**Policy Insights**

1. In many coal-burning areas, the decline of coal is slowed down by long-term contracts that insulate coal-fired generation from competition.
2. Compensation for early closure is a "pay-to-break" mechanism which may be a politically feasible alternative to "polluter-pays" policies such as carbon pricing.
3. Competitive auctions can deliver efficient and transparent compensation payments for early coal closure relative to negotiations which suffer from asymmetric information.
4. Ensuring that there is sufficient participation in a coal closure auction through design adjustments can ensure discovery of true closure costs.
5. Shutting down coal-fired power plants can lead to fiscal savings in countries where coal-fired generation is supported by capacity payments.

# 1. Introduction

Fossil fuels are the world's greatest source of greenhouse gas emissions. There is already enough fossil fuel energy in the system that if these assets are operated until the end of their economic lifetimes, the global average temperature is expected to rise above 1.5°C thereby breaching the goals of the Paris Agreement (Tong et al. 2019). Coal combustion accounts for 40% of annual emissions from energy and industry, and is the most carbon-intensive form of energy generation (Andrews and Peters, 2022). Due to the unproven nature of carbon capture and storage at scale and the credibility crisis in the carbon offset market (Bacilieri et al. 2023; Cullenward and Victor 2020), many countries have consequently put in place a coal phaseout target (Fig 1).

**Figure 1: Coal phaseouts**

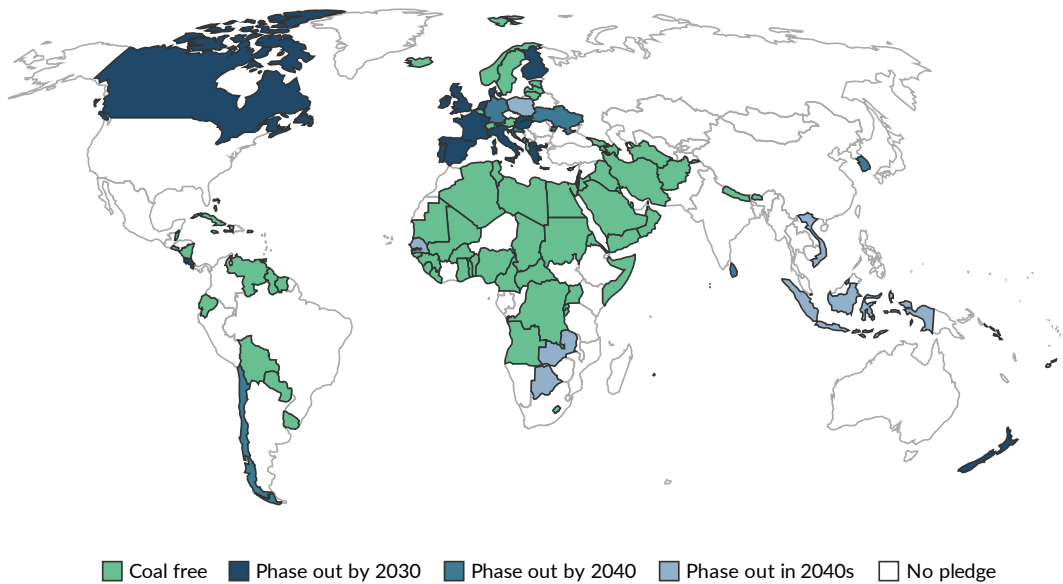

Data source: Powering Past Coal Alliance; Ember Climate; Beyond Coal EU; Bloomberg Coal Countdown and other sources
Note: Where a concrete phase out date is not defined, we have allocated the final year of the target decade. For example, "Phase out in the 2040s" is given a target date of 2049.
OurWorldInData.org/energy | CC BY

Phaseout policies need to restrict the growth of new fossil fuels and expedite the retirement of the existing stock. While there is rich scholarship on supply-side policies to achieve the former (e.g., Harstad 2012, Lazarus and van Asselt 2018), there is relatively less on the latter. Recent work has discussed the complex mechanics of closing Chinese-funded coal-fired power plants abroad (Clark et al 2023). Yet analysis on the efficiency of different exit policies or their susceptibility to rent-seeking due to information asymmetries is still limited. Premature exit is complicated as it requires breaking existing long-term contracts, conflicts with investor protection norms and may cause issues with security of electricity supply. Compensation for early closures is a "pay-to-break" mechanism that accounts for the costs of prematurely winding down coal-fired power plants.

Closing down coal-fired power plants involves various costs: severance packages need to be paid to workers, early termination fees are required if contracts are broken prematurely, and polluted waterways or soil need rehabilitation. Furthermore, in cases where power plant closure is not due to market forces, then there also needs to be a payment to reflect the opportunity cost of early closure, that is, the net present value of the future stream of profits that would have otherwise occurred. The opportunity cost can vary significantly by power plants based on where they are located, whether they supply to the grid or are engaged in captive generation, their efficiency, and the extent of competition from other sources of energy.

It is difficult for a policymaker to know, ex-ante, what the opportunity cost of closure is, and it is this informational challenge, that makes auctions for compensation a useful policy. Owners of coal-fired power plants have an incentive to overstate closure costs while governments, considering taxpayer interests, seek to achieve low-cost closures. In bilateral negotiations, the power plant owner can exploit this information asymmetry. By contrast, an auction with enough competition can lead to the discovery of true closure costs because if a coal plant owner overstates costs, they risk losing the auction and exiting with no compensation, and if they understate costs, they similarly incur losses.

### 1.1. *When is a coal phaseout economically rational?*

A coal phaseout target can be theoretically rationalized if the wedge between social and private costs is so large that the socially optimum quantity of coal is zero (Fig. 2). Apart from the climate externality, coal combustion contributes to soil and water pollution, sulphur dioxide emissions, fine particulate matter, habitat degradation, and negative health impacts (Barrows et al. 2019). As evidence accumulates around the size of these externalities, the erstwhile cost-benefit calculations that justified continued coal combustion, may no longer hold up, at least in certain locations (Rafaty et al. 2020).

While addressing coal combustion's various market failures separately is likely to be more efficient because there could be distinct responses to each market failure, such as locating away from high value habitats (to address the biodiversity externality) or installing pollution control equipment (to address negative health impacts), in practice, each policy is administratively costly to implement and there are hurdles towards a credible system of Pigouvian taxation or even Coasian bargaining for natural capital (Teytelboym 2019; Deryugina et al. 2021). There are many examples of substances that were deemed to be net beneficial in the past but which were subsequently phased out due to growing evidence of harm or diminished benefits due to the emergence of cost-competitive substitutes (e.g. asbestos).

**Figure 2: The economic rationale for a coal phaseout**

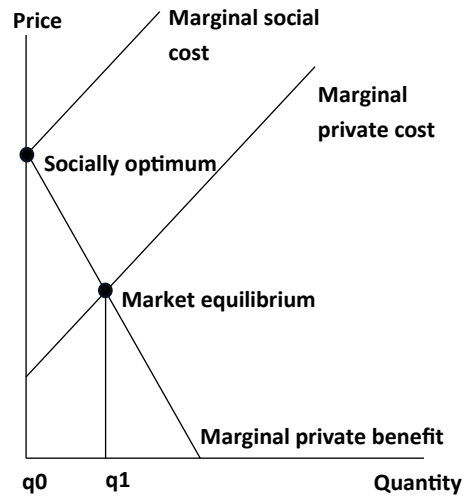

The mean value of the social cost of carbon from expert analyses is now $500/tCO$_2$ (Moore et al. 2023) and there is not a single carbon pricing scheme in the world that currently comes close to this value (World Bank 2023). There is also, increasingly, a re-orientation by economists to find the most *efficient* way to reach targets set by political or scientific consensus rather than second-guess these targets themselves. For example, most economists have now ceased to question the level of optimal warming, and instead, focus on how to keep global average temperatures below what is deemed as unsafe by scientists. On coal combustion, which claims more lives than any other source of energy (Ritchie 2020), one could similarly argue that economists may increasingly start to think about how to ration the remaining coal towards the uses with the highest value, without second guessing the "optimal" quantity of coal in the system.

*1.2.    Comparison with carbon pricing*

A coal phaseout is a technology-specific policy. This contrasts with a carbon price which is technology-neutral since it does not tell agents how to cut down on emissions (they are free to replace fossil fuels with renewable energy, increase energy efficiency, or stop a certain activity altogether). A carbon price is theoretically first best as it can lead to the discovery of least cost abatement strategies thanks to its non-prescriptive nature and singular focus on penalizing the pollutant.

However, in much of the developing world, including major coal burning nations such as India, Indonesia and South Africa, the carbon price signal will not work as expected because of the lack of competitive markets. In these jurisdictions, a utility procures electricity from independent power producers (IPPs) using long-term contracts spanning 2-3 decades. These contracts make carbon pricing far less effective because IPPs can simply pass on the full carbon cost to the utility (if the contract is cost-plus) or the IPP is not exposed to market dynamics at all (because the contract is fixed price). In India, there is documented variation in how IPPs react to a carbon price, with those under long-term contracts being far less reactive than those operating on the wholesale market (Khanna 2024). Renewable energy support policies (such as subsidies) also have to contend with same market rigidities. In these jurisdictions, explicitly mandating the exit of coal by a certain date and then getting coal-fired power plants to bid for scarce competition can create competition in the absence of the market itself being dynamic and competitive. Closure is also important since coal-fired power plants are supported by power purchase agreements that require annual capacity payments even if no electricity is purchased from the plant (Kansal et al. 2023, Srivastav et al. 2024). A pay-to-break mechanism can, therefore, represent important fiscal savings.

In more advanced economies where electricity is procured through intra-day or day-ahead auctions, and where dispatch is governed by a ranking of marginal costs (the "merit order"), the bids of polluting generators would theoretically reflect the carbon cost. This would affect coal's ranking in the merit order, potentially pushing it out of use or at least reducing the quantity combusted. The higher the carbon price is, the greater the chance that coal is not used to meet electricity demand. Yet, experts regularly assert that the carbon price is still far too low or not stable enough to create the incentives required for the low-carbon transition at scale (World Bank 2023).[1] According to the World Bank, "as of April 2023, less than 5% of global greenhouse gas emissions are covered by a direct carbon price or are at or above the range recommended by the High-Level Commission on Carbon Prices (USD 40 – 80 per tCO2)" (World Bank 2023).[2,3]

Due to the reality of being in a second-best world, where carbon pricing is inherently constrained to be below the social cost of carbon or where it simply cannot pass through to affect the decisions of market agents, there is an opportunity space to consider multiple policy instruments to achieve climate goals (Lipsey and Lancaster 1956; Jenkins 2014), an idea which gains further traction when one considers that there are also innovation and credit-related market failures that affect the transition to clean energy (Jaffe et al 2005). Therefore, across jurisdictions, we simultaneously see carbon pricing alongside technology-specific measures such as fuel efficiency mandates for cars, gasoline taxes, building regulations, clean innovation subsidies, and feed-in tariffs for renewable energy (Jenkins 2014; Vogt-Schilb and Hallegatte 2017).

### *1.3.   Political economy and feasibility*

There are also pragmatic reasons why technology-specific policies are implemented. If there is no measurement infrastructure to track emissions, a feasible alternative is to target a close proxy such as coal combustion, which is easier to measure. Even in jurisdictions with measurement infrastructure, explicitly tracking fossil fuel infrastructure like coal-fired power plants and combustion volumes is a transparent way to create greater accountability while focusing only on net emissions requires building up sophisticated carbon accounting systems, which are at risk of regulatory capture and lobbying by vested interests if not done well (Green and Kuch 2022). This risk is non-trivial as illustrated by the unravelling of the carbon offset market which has large volumes of offsets that have questionable claims of additionality (Cullenward and Victor 2020).

---

[1] Furthermore, in many markets where there is no merit order dispatch and the fixed costs of the power plant have been paid off. Here, as long as the revenue minus the fuel cost and the carbon cost is non-negative, the power plant will continue to operate.

[2] While some carbon pricing schemes have led to a substantial decline of coal (see the case of the United Kingdom discussed in Leroutier 2022), the emphasis here is on achieving a decline that is commensurate with fully closing the wedge between the private and social costs of coal combustion.

[3] Even in jurisdictions with a high carbon price, coal closures may not happen at the expected rate because of capacity markets, which pay for back-up generation in case there are unexpected spikes in demand. There was a concern that capacity markets were biased towards fossil fuels and that demand side management was under-utilised to manage the grid even though it was the "lower carbon" solution (Creutzig et al. 2018, Vaughan 2018).

There is also a difference between punitive measures such as carbon pricing and compensation for early closure. While the former generates government revenues, the latter requires expenditure from the government. While the former generates resistance from industry, the latter is argued to be more palatable, making it a pathway towards climate action in contexts where there is a well-mobilised fossil fuel lobby (Srivastav and Rafaty 2023). Furthermore, since coal phaseouts relate to discrete and highly salient outcomes that are more tangible than outcomes such as reducing $CO_2$eq/GDP (Collier and Venables 2014, Erickson et al 2018), they might be useful for virtue signaling amongst a climate-conscious electorate or in international climate forums where high-income countries are expected to contribute relatively more to climate change mitigation.[4] Questioning the social license to operate polluting assets such as pipelines, mines or powerplants has already seen significant levels of political mobilization around the world (Piggot 2018).

Notwithstanding the various arguments in favour or against a technology-specific policy, this piece takes as a given the science-based imperative to phase out unabated coal combustion and focuses on how compensation for early closure can be implemented in a manner that is efficient. We explore when an auction-based approach is appropriate, how to ensure additionality, appropriately manage dynamic incentives, and think through system-wide effects.

## 2. Auctions, negotiations, and other incentives: when to use what

Which type of coal phaseout policy to implement depends on the context and a variety of policies have been deployed across jurisdictions (Table 1). If there is no auction design that can stimulate enough competition, which might be the case in highly concentrated markets,[5] measures such as strengthened incentives for scrappage or repurposing of coal assets (e.g., by switching to biomass), removal of fossil fuel subsidies, and cap-and-trade schemes to restrict tonnage of coal combusted may be better suited, if there is a political will to expedite the decline of coal. For example, in Chile, where there was concentrated coal plant ownership, an incentive scheme to scrap the coal asset and in return, get a loan to build out renewable energy infrastructure was used (Climate Investment Funds 2021).

A negotiated approach to compensation payments for early retirement could also be considered but this would require managing the information asymmetry. Alberta, which in 2015 decided to phaseout coal, had to set aside a little over 1 billion Canadian dollars to compensate coal power companies (Vriens 2018). This involved negotiations with three major private coal plant owners.

---

[4] High income nations have already begun establishing vehicles to fund for plant closures in developing nations. An example is the "Just Energy Transition Partnership" in Indonesia in which "20 billion USD [will be mobilised] …to help phase out coal energy and invest in renewable energy infrastructure" (UK Cabinet Office 2022).
[5] Careful analysis is needed to determine if a competitive auction is truly infeasible. Often by adjusting geographic scope, participation rules, and timing, competition can be created.

Where coal-fired power plants are publicly owned, and there is an impetus to internalise the social cost of pollution, the relevant public authority can simply optimise like a social planner and there is no need for market-based instruments (there is no principal-agent problem or informational asymmetry, and simple mandates and moratoriums can be deployed) (Benoit et al 2022). For example, in Ontario, where coal went from constituting a quarter of total supply in the early 2000s to 0% by 2014, the phaseout strategy had a command-and-control nature since all of the remaining coal-fired power plants were owned by a provincial body, Ontario Power Generation (OPG). OPG wrote off whatever residual capital value remained in the power plants to help facilitate closure, while the government and grid operator managed the power system through the installation of new natural gas and nuclear power plants (Winfield and Saherwala 2022). The plan for a total phaseout was motivated by concerns around smog, acid rain, and heavy metal pollution (Winfield and Saherwala 2022).

In markets where there is private ownership and enough diversity in coal plant ownership, auctions can be used to close down coal-fired power plants. It is often possible to adjust the design of the closure auction to ensure sufficient competition (e.g. by calibrating the geographic scope, adjusting the eligibility criteria and deciding if the auction is broken into multiple rounds or not). The German Coal Exit Act (2020) allowed hard coal power plants to voluntarily exit between 2020 and 2026 and compete in auctions for compensation payments (Tiedemann and Müller-Hansen 2023). It stated that from 2027, decommissioning could be mandated without compensation thereby incentivising participation, and by 2038, all coal plants needed to retire. Coal power plants submit sealed bids stating their compensation amounts in €/MW which are ranked in descending order. The most cost-effective bids are selected until the total budget for compensation payments is exhausted.

Since Germany is the only country in the world till date to have implemented a coal phaseout auction, we use it to analyse the merits and demerits of the policy, with an acknowledgement that more evidence is likely needed to establish the feasibility of such a scheme and not all lessons will be transferable across contexts.

Table 1: Coal phaseout strategies in select jurisdictions

| Region | Privately-owned coal companies | Strategy | Rationale |
|---|---|---|---|
| **Ontario** | ✗ | Write-off value | Aligned incentives since coal plants publicly owned |
| **Alberta** | ✓ | Negotiated compensation | Private players, insufficient competition |
| **Germany (hard coal)** | ✓ | Auction-based compensation | Private players but sufficient competition |
| **Germany (lignite)** | ✓ | Negotiated compensation | Private players, insufficient competition |
| **Chile** | ✓ | Incentive to scrap coal and re-invest in renewable energy | Private players, insufficient competition |

## 3. Ensuring additionality

A challenge for any coal closure scheme is to ensure *additional* emission savings which would not have occurred otherwise, or which are not offset by higher emissions elsewhere. In a negotiated approach, it is difficult to ascertain if the coal power plants asking for compensation were going to wind down anyway. Due to cheap renewable energy, many coal power plants are becoming economically unviable due to low utilization rates. In Australia, this is creating economic pressure to retire coal-fired power plants ahead of schedule (Jotzo et al 2018), as evidenced by a plant shutting down a decade ahead of its scheduled closure (Turnbull 2022).[6] A competitive auction can theoretically reflect the opportunity cost of closure by getting lower bids from power plants with shorter economic lifetimes. However, to ensure this happens in practice, a few factors require consideration.

---

[6] In the Australian context, since the power market is liberalised and competitive, taxing carbon could expedite the decline of coal. Yet, the political economy of Australia makes this type of policy difficult to implement and sustain, as evidenced by their scrappage of a carbon tax only two years after it was introduced. A coal closure auction could, by contrast, appease a well-mobilised lobby while still delivering on climate goals.

Winding down operations can take a long time and there is a risk that if the lead time for decommissioning is too short, only those bidders who have already planned to mothball or decommission their plant participate in the auction. In the first German auction round in 2020, the lead time was only one month from the date of award until the end of operations, raising concerns that payments were given to power plants which were already planning for closure.

A second concern is that the retirement of coal can interact negatively with other climate policies. In Germany, coal retirement carries the risk of creating a "waterbed effect" in the European Union Emissions Trading Scheme (EU ETS). Coal-fired power plants demand permits to emit $CO_2$ (EUAs) but the gap in demand created by their closure at a constant supply of EUAs would cause the EUA price to fall, leading temporarily to increased emissions elsewhere and diluting the additionality of the coal closure scheme (Akerboom et al 2020).[7] To address this, the German government announced it would cancel an equivalent amount of EUAs.[8] Overall, however, given the nature of the EU ETS with its absolute emissions cap, total emissions would not have increased, even if the price were to fall.

There is also a risk of international leakage, where restricting coal combustion in one area may result in an increase in coal combustion in other areas that is not under similarly stringent climate policy (see Fig 3). This leakage can be offset by restricting coal supply (Asheim et al. 2019). A carbon border adjustment mechanism can achieve a similar effect by creating an incentive for non-complying areas to implement more stringent climate policy (Helm et al. 2012). However, in many cases, coal markets are not integrated, ruling out the possibility of such leakage (there is no global supply curve; only local supply curves). For example, German lignite and Indian coal have sufficiently low energy densities such that it is uneconomic to transport these resources large distances.

**Figure 3: The risk of leakage**

---

[7] Assuming German coal is significant enough to affect prices in the EU ETS.
[8] Based on modelling around how many carbon permits these power plants would have demanded.

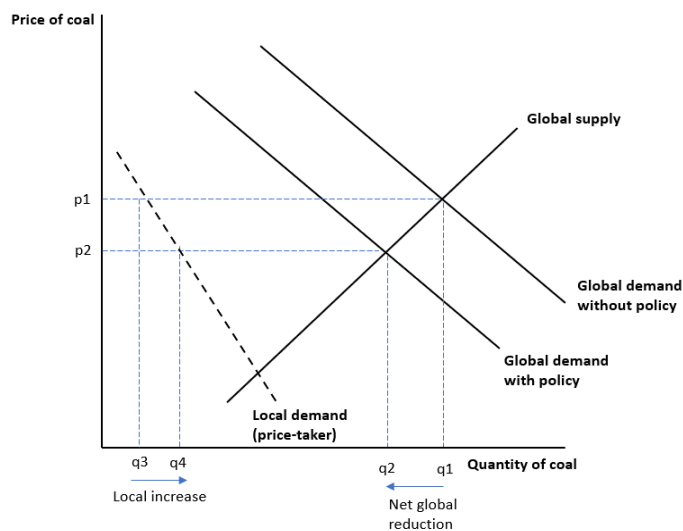

Note: demand-side policy can reduce global demand for coal (q1 → q2) but in price-taking jurisdictions with no policy, the resultant fall in global price (p1 → p2) can increase local consumption thereby resulting in "leakage" (q3 → q4). The overall fall in global demand would be higher if these jurisdictions also adopted climate policy. Shifting global supply to the left can help counter leakage and achieve higher net global reduction.

## 4. Dynamic incentives

It is also important to consider dynamic incentives. Germany's coal phaseout auction is broken into multiple rounds over time. While a staggered phaseout has the advantage of giving the system time to adapt, it also makes each round less competitive because only a sub-set of coal-fired power plants participate. It may also create strategic considerations amongst bidders around which auction round to participate in. For example, if electricity demand is growing and renewable expansion cannot keep apace, then expected profits for the last coal power plants are higher. Coal power plants will internalize and reflect this via bids. They may also delay participation. However, too much strategic waiting can create a lack of competition that undermines the efficiency of earlier auction rounds.

In this situation, to ensure enough competition in earlier rounds, auction rules can penalize late-movers. Three ways of doing this from the German case include (i) decreasing the maximum possible compensation amount over time, (ii) instating an end rule such that closures after a certain date are mandated without compensation, and (iii) imposing penalties for lack of participation (if there is undersubscription in a German auction round, the oldest coal power plant must close down without compensation).

Furthermore, power plants remaining in the market could be mandated to fund the compensation for early closure in proportion to their $CO_2$ emissions (Jotzo and Mazouz 2015). This acts as a penalty for staying longer in the market and adheres to the polluter pays principle. There is a risk that since it creates costs for coal investors, they may challenge the policy in court. This happened in the case of Germany's nuclear phaseout where the government was straddled with several costly litigation cases afterwards and was also the case in Alberta, which is one of the reasons why the government had to compensate its coal companies (Winfield and Saherwala 2022).

A potentially more sophisticated and flexible approach would be to allow for menu bids that let coal plants signal individual decommissioning costs for different decommissioning dates through their bids (Riechmann and Zaehringer 2020). In this case, the auctioneer invites bids for different decommissioning dates for a certain deadline, for example, in 2020 for closure in 2021 to 2028, aligned with the target path for the coal phaseout. This would allow plants with long lead times (such as coal plants with heat output) to also compete, thereby bolstering participation. This can help determine combinations of closure dates that fulfill the target path and minimize total compensation payments. It also maintains a phased approach that is more conducive to system stability.

Finally, there is a concern amongst some that compensation creates moral hazard, where if it is known that there will be payments for early closure, coal-fired power plants will enter the market opportunistically. However, a new power plant will need relatively high compensation relative to existing power plants, making it more likely to lose the auction for compensation payments. It should also be noted that in most jurisdictions, there is no economic case for new coal, while in others, new coal plant construction is simply prohibited by law (such as in Germany).

## 5. System-wide effects

The early closure of coal plants can have system-wide effects. For example, some power plants may have positive externalities because they help manage congestion while others may help in balancing the grid. These externalities will not show up in individual bids as they relate to the system rather than the individual coal power plant. This can lead to situations where assets are inefficiently marked for closure because their system value has been ignored. Design adjustments in the auction can ensure that critical assets are not decommissioned prematurely. In Germany, southern coal-fired power plants were not allowed to participate in the first auction due to system stability concerns. Subsequently, when these were allowed to participate, a "penalty" was applied on top of their bid to reflect the higher system costs of their closure. The German power regulator could also keep these power plants as reserve capacity.

The auction design could also incorporate some flexibility given the potential for crises to occur in the power market. The progressive decline in German bid caps (which specify maximum compensation) turned out to be incompatible with the high levels of scarcity which prevailed after the Russian invasion of Ukraine (Riechmann et al. 2023). The 2020 bid caps prevented these scarcity signals from being reflected via higher bids for coal closure, thereby impeding the ability of the auction to deliver efficient decisions (Riechmann and Zaehringer 2020, Riechmann et al. 2023). The German government had to retrospectively allow plants that had stopped operations to re-enter the market. As of 9th January 2023, there are 12 power plants, with a total capacity of 4,166 MW (40% of total procured capacity) that have declared their temporary return to market. This crisis was extreme. However, it underscores the importance of maintaining some degree of flexibility and complementing coal phaseout programmes with complementary policies that seek to expand clean energy generation to ensure security of supply.

## 6. What to auction

There is also a question of *what* to auction. In Germany, the auction considered coal capacity rather than generation, leading to the result that in first auction round two-thirds of the capacity slated for retirement belong to modern coal plants rather than older ones (Riechmann & Zaehringer 2020). Bids in Germany are re-ranked by carbon emissions, and since modern plants are run more often - owing to their lower costs and higher efficiency - their emissions were naturally higher. However, if emissions were scaled by generation, as is appropriate to gauge efficiency accurately, it would have become clear that older plants represent worse value when it comes to emissions per unit generation. This illustrates the dangers of auctioning off the wrong variable and re-ranking bids inappropriately.

Apart from generation, auctions could also consider tonnage of coal combusted, which is an even closer proxy to avoided emissions. The government could set a cap on the remaining tonnage of coal which can be burnt, in line with the net zero pathway, and permits representing tonnage of coal could be auctioned off. This scheme, in contrast to the earlier proposal, would follow the polluter pays principle where coal-fired power plants must pay for permits rather than get compensated for retiring entire power plants. This is useful in settings where governments are budget constrained and would benefit from the tax revenues.[9] A secondary market could allow for the trading of these permits, where more efficient coal-fired power plants would be able to reduce coal input and sell permits, while less efficient ones would have to incur the costs of buying these (Riechmann et al. 2023). The least efficient coal-fired power plants would have to exit because costs become too high to justify operations. If there is large variance in the grades of coal used, the auction could also establish an "exchange rate" between permits relating to different grades of coal.

There is 30 years of experience with emissions trading systems word-wide which shows that cap-and-trade systems, if designed well and appropriately implemented, can achieve targeted emissions reductions cost-effectively (Schmalensee & Stavins 2017). A cap-and-trade market for tonnage of coal might be easier to design than the far more complex emission trading schemes such as the EU ETS. For example, in developing countries where measurement of $CO_2$ emissions is difficult due to limited resources and state capacity, tonnage of coal may represent a more practical option. However, there may be distributional effects: if allowances are grandfathered rather than competitively auctioned, then the largest polluters get relatively more allowances which may create distortions if the market is not liquid enough.

**Conclusion**

Many parts of the world still do not have any form of carbon pricing and out of those that do, no jurisdiction has a price that comes close to the latest estimates of the social cost of carbon (World Bank 2023, Moore et al. 2023). Because of this, coal combustion and other polluting activities are not falling as fast as they should to internalise the climate externality.

---

[9] Targeting tonnage of coal is an "intensive margin" strategy while retiring power plants is an "extensive margin" policy. The latter can help break out of capacity payments i.e., lump sum payments that have to be made to a power plant even if it is selling zero units of electricity on the market. Capacity payments are common across many power markets and constitute a large share of power market debt (Srivastav et al. 2024).

Coal, in particular, is the most polluting form of energy on the planet and stabilising global average temperatures requires critically managing coal combustion. Due to the credibility crisis of the carbon offset market (Cullenward and Victor 2020) and the unproven nature of carbon capture and storage at scale (Bacilieri et al. 2023), the maths of net zero by 2050 necessitates a quantity reduction in coal combustion. This means no new coal development and the early retirement of existing coal-fired power plants (Tong et al. 2019). While there is a rich literature on supply-side policy to inhibit the development of new coal infrastructure (Harstad 2012, Lazarus and van Asselt 2018), there is relatively little scholarship on how to *efficiently* wind down operational coal plants before the end of their economic lifetime. This is the gap our article addresses.

Many coal-fired power plants in the world are insulated from competition by renewable energy because they are under long-term power purchase agreements (Kansal et al. 2023). In such settings, removing barriers to renewable energy deployment is not be enough. Policies such as compensation for early closure represent a "pay-to-break" mechanism that can facilitate contractual exit and make room for cheaper and cleaner alternatives.

However, how much should be paid for early exit and in what order should coal plants be retired? Negotiated buyouts suffer from asymmetric information while auctions can reveal true opportunity cost of closure as well as the efficient order of exit as long as there is no collusion between bidders. In markets where coal-fired power plants are privately owned by a diverse group of firms, then auctions offer a way to address informational asymmetries and should be designed in a manner that is sensitive to concerns around additionality, dynamic incentives, and system-wide effects to obtain best results. Where there is concentrated ownership, strengthened incentives for scrappage may be better suited.

Further, from a political-economy perspective, compensation for early closure may be expedient. The owners of coal-fired power plant represent a highly mobilised political group that can block progress unless carefully managed (Stokes 2020). In Germany, the coal phaseout scheme also earmarked funds for workers and coal-dependent regions to help in reskilling, re-development, and the rehabilitation of polluted lands. Unless coal-dependent communities are compensated in the wake of large power plant closures, there is a risk they may get politically polarised and marginalised (Heal and Barry 2017).

Finally, an emergent reality is that analyses on the "optimal" quantity of coal may quickly be replaced by conversations on *how* the world can reach coal phaseout targets in a manner that reflects the opportunity cost of early retirement, minimises rent-seeking, and which explicitly prices in rehabilitation to workers and degraded lands to ensure a just transition. This is similar to how economists no longer second-guess the optimality limiting global temperature rise to 1.5°C (Stern et al. 2022), and instead have shifted the focus on policies to achieve these aims.

## Acknowledgements

SS acknowledges support from the Climate Compatible Growth programme.

Both authors thank Moritz Schwarz, Matthew Ives, Sam Fankhauser and Alex Clark for feedback.

## Author information

1. Smith School of Enterprise & the Environment, Department of Economics, Institute of New Economic Thinking, University of Oxford, United Kingdom

   Sugandha Srivastav

2. Frontier Economics, Cologne, Germany

   Michael Zaehringer

Correspondence to Sugandha Srivastav (sugandha.srivastav@smithschool.ox.ac.uk)

## Ethics declaration

The authors declare no competing interests.
Michael Zaehringer works at Frontier Economics but views expressed are of the individual and do not reflect that of Frontier Economics in any form.